\documentclass[onecolumn,amsmath,amssymb]{qm2008_abs}
\usepackage{graphicx}
\usepackage{dcolumn}
\usepackage{bm}
\begin{document}

\title{\large Radiative and Collisional Jet Energy Loss in a Quark-Gluon Plasma}

\bigskip
\bigskip
 
\author{ G.Y.~Qin$^1$} \email{qing@physics.mcgill.ca}

\author{ J.~Ruppert$^{1,3}$}
\author{ C.~Gale$^1$}
\author{ S.~Jeon$^1$}
\author{ G.~D.~Moore$^1$}

\author{ M.~G.~Mustafa$^2$}
 
\affiliation{$^1$ Department of Physics, McGill University, Montreal, QC, H3A 2T8, Canada \\ $^2$ Saha Institute of
Nuclear Physics, 1/AF Bidhannagar, Kolkata, 700064, India \\ $^3$ Institut f\"ur Theoretische Physik, J.W.~Goethe University
Frankfurt, Max-von-Laue-Str.~1, D-60438 Frankfurt am Main, Germany}

\bigskip
\bigskip

\begin{abstract}
\leftskip1.0cm
\rightskip1.0cm

We calculate radiative and collisional energy loss of hard partons traversing the quark-gluon plasma created at RHIC and
compare the respective size of these contributions. We employ the AMY formalism for radiative energy loss and include
additionally energy  loss by elastic collisions. Our treatment of both processes is complete at leading order in the
coupling, and accounts for the probabilistic nature of jet energy loss. We find that a solution of the Fokker-Planck
equation for the probability density distributions of partons is necessary for a complete calculation of the nuclear
modification factor $R_{AA}$ for pion production in heavy ion collisions. It is found that the magnitude of $R_{AA}$
is sensitive to the inclusion of both collisional and radiative energy loss, while the average energy is less
affected by the addition of collisional contributions. We present a calculation of $R_{AA}$ for $\pi^0$ at RHIC,
combining our energy loss formalism with a relativistic (3+1)-dimensional hydrodynamic description of the thermalized
medium.

\end{abstract}
 
\maketitle

\section{Introduction}

The main goal of relativistic heavy ion collisions is to study strongly interacting nuclear matter at
extreme temperatures and densities and investigate the possibility of a phase transition between hadronic matter and
a quark-gluon plasma (QGP). Experiments at the Relativistic Heavy Ion Collider (RHIC) have demonstrated a significant
suppression of high $p_T$ hadrons produced in central $A+A$ collisions in comparison to those in binary scaled $p+p$
collisions \cite{Adcox:2001jp, Adler:2002xw}. This result has been referred to as jet quenching and has been
attributed to the energy loss of hard $p_T$ partons in a hot quark-gluon plasma \cite{Gyulassy:1993hr}.
Experimentally, jet quenching can be characterized by measurements of various quantities, i.e., the nuclear
modification factor $R_{AA}$, the elliptic flow $v_2$ at high $p_T$, and high $p_T$ hadron correlations, etc.

In \cite{Qin:2007zz} we presented a systematic calculation of $R_{AA}$ for $\pi^0$ in central and non-central Au+Au
collisions at $\sqrt{s} = 200$~AGeV combining the Arnold, Moore and Yaffe (AMY) \cite{Arnold:2001ms} formalism with a
(3+1)-dimensional ideal hydrodynamical model \cite{Nonaka:2006yn}. In this contribution, we consistently incorporate both
collisional and radiative energy loss in the same framework \cite{Qin:2007rn}.

\section{Jet Energy Loss in the AMY Formalism}

In the AMY formalism, the evolution of the jet momentum distribution $P(E,t)={dN(E,t)}/{dE}$ in the QGP is obtained by solving
a set of coupled Fokker-Planck type rate equations for quarks plus anti-quarks and gluons, which have the following
generic form,
\begin{equation}
\label{jet-evolution-eq} \frac{dP(E)}{dt} = \int_{-\infty}^{\infty} d\omega \left[P(E{+}\omega)
\frac{d\Gamma(E{+}\omega,\omega)}{d\omega dt} - P(E)\frac{d\Gamma(E,\omega)}{d\omega dt}\right]
 \end{equation}
where ${d\Gamma(E,\omega)}/{d\omega dt}$ is the transition rate for processes in which partons of energy $E$ lose
energy $\omega$. The $\omega < 0$ part of the integration incorporates processes which increase a particle's energy.
The radiative part of the transition rates is discussed in
\cite{Jeon:2003gi, Turbide:2005fk, Qin:2007zz}. In the following, we address the calculation of elastic collision
rates \cite{Qin:2007rn}.

As collisional energy losses are more sensitive to small energy transfers in comparison to radiative ones, it should be an
adequate procedure to
approximate elastic collisions by a mean energy loss plus a momentum diffusion term as dictated
by detailed balance. If elastic collisions turn out to dominate jet quenching we may want to improve this treatment,
but if they are subdominant it should be sufficient to quantify their effect.

As we perform the numerical computation of discretized Eq.~(\ref{jet-evolution-eq}) such that $\int d\omega
\rightarrow \Delta_\omega \sum_{\omega = n \Delta_\omega}$, the transition rates for elastic collisions are
approximated by two spikes at $\omega=\pm \Delta_\omega$,
\begin{eqnarray}
\frac{d\Gamma(E,\omega)}{d\omega dt} \!\!&\approx&\!\! \left[1+f_B(\Delta_\omega)\right]\frac{1}{\Delta_\omega}
\frac{dE}{dt} \delta(\omega-\Delta_\omega) + f_B(\Delta_\omega)\frac{1}{\Delta_\omega} \frac{dE}{dt}
\delta(\omega+\Delta_\omega)\, ,\ \ \ \ \ \ \
\end{eqnarray}
with $f_B$ the Bose-Einstein thermal distribution function. For small $\Delta_\omega$, the above collisional
transition rates yield the correct energy loss rate and preserve the detailed balance. In the small $\delta_\omega$
limit, this
procedure to incorporate collisional energy loss is equivalent to introducing the drag term $(dE/dt)dP(E)/dE$, and the
diffusion term $T(dE/dt)d^2P(E)/dE^2$ into Eq.~(\ref{jet-evolution-eq}).

At leading order, the mean energy loss by elastic collisions is calculated in kinetic theory,
\begin{eqnarray}
\label{dEdt} \frac{dE}{dt} \!\!&=&\!\! C_{\rm coll} \pi \alpha_s^2 T^2 \left[\ln\frac{ET}{m_g^2} + D_{\rm coll}\right]
\end{eqnarray}
where $C_{\rm coll}$ and $D_{\rm coll}$ are ${\cal O}(1)$ constants depending on different channels \cite{Qin:2007rn},
and $m_g^2 = 2 \pi \alpha_s T^2 (1 + N_f/6)$ is the thermal gluon mass.

\section{Results at RHIC}

Now we employ this formalism to a realistic description of energy loss of hard $p_T$ leading partons in the soft
nuclear medium in Au+Au collisions at $\sqrt{s} = 200~{\rm AGeV}$ at RHIC. We utilize a fully (3+1)-dimensional
hydrodynamical model for the description of the thermalized medium created in the collisions as it has been shown to
give a good description of bulk properties at RHIC \cite{Nonaka:2006yn}. The product of the initial hard parton
densities is determined from the overlap geometry between two nuclei in the transverse plane of the collision
zone. The initial momentum distribution of hard jets is computed from perturbative QCD, using the factorization theorem.
The evolution of
jet momentum distribution in the thermalized medium is evaluated in AMY \cite{Arnold:2001ms}, solving
Eq.~(\ref{jet-evolution-eq}) with both collisional and radiative energy loss. The final hadron spectrum at
high $p_T$ is obtained by the fragmentation of jets in the vacuum after their passing through the (3+1)-dimensional
expanding medium. For further details see \cite{Qin:2007zz} where the radiative energy loss has been studied in an
analogous manner.

The nuclear modification factor $R_{AA}$ is defined as the ratio of the hadron yield in A+A collisions to that in p+p
interactions scaled by the number of binary collisions $N_{\rm coll}$:
\begin{eqnarray}
R^h_{ AA}(b,\vec{p}_T,y) &=& \frac{1}{N_{\rm coll}(b)} \frac{{dN^h}(b)/{d^2p_Tdy}} {{dN^h_{ pp}}/{d^2p_Tdy}}.
\end{eqnarray}

In the AMY formalism, the strong coupling constant $\alpha_s$ is the only quantity which is not uniquely determined by the
model, once the temperature and flow profiles are fixed by the initial conditions and subsequent evolution of the
(3+1)-dimensional hydrodynamics. In this work, we take its value to be constant at $\alpha_{\rm s}=0.27$, which
reproduces the most central data (see Fig. \ref{fig34}).

\begin{figure}[htb]
\begin{center}
\includegraphics[width=7.4cm]{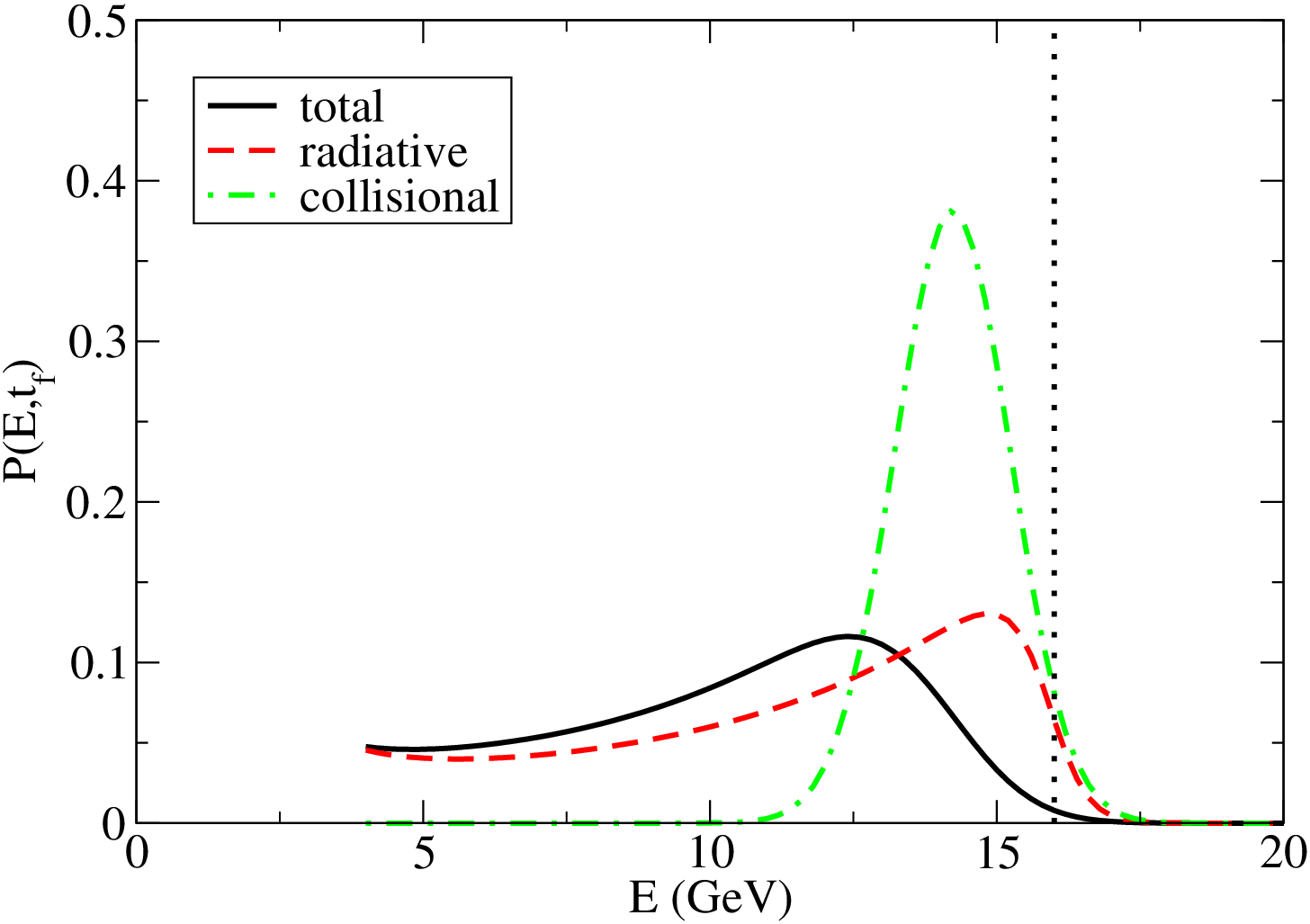}
\includegraphics[width=7.4cm]{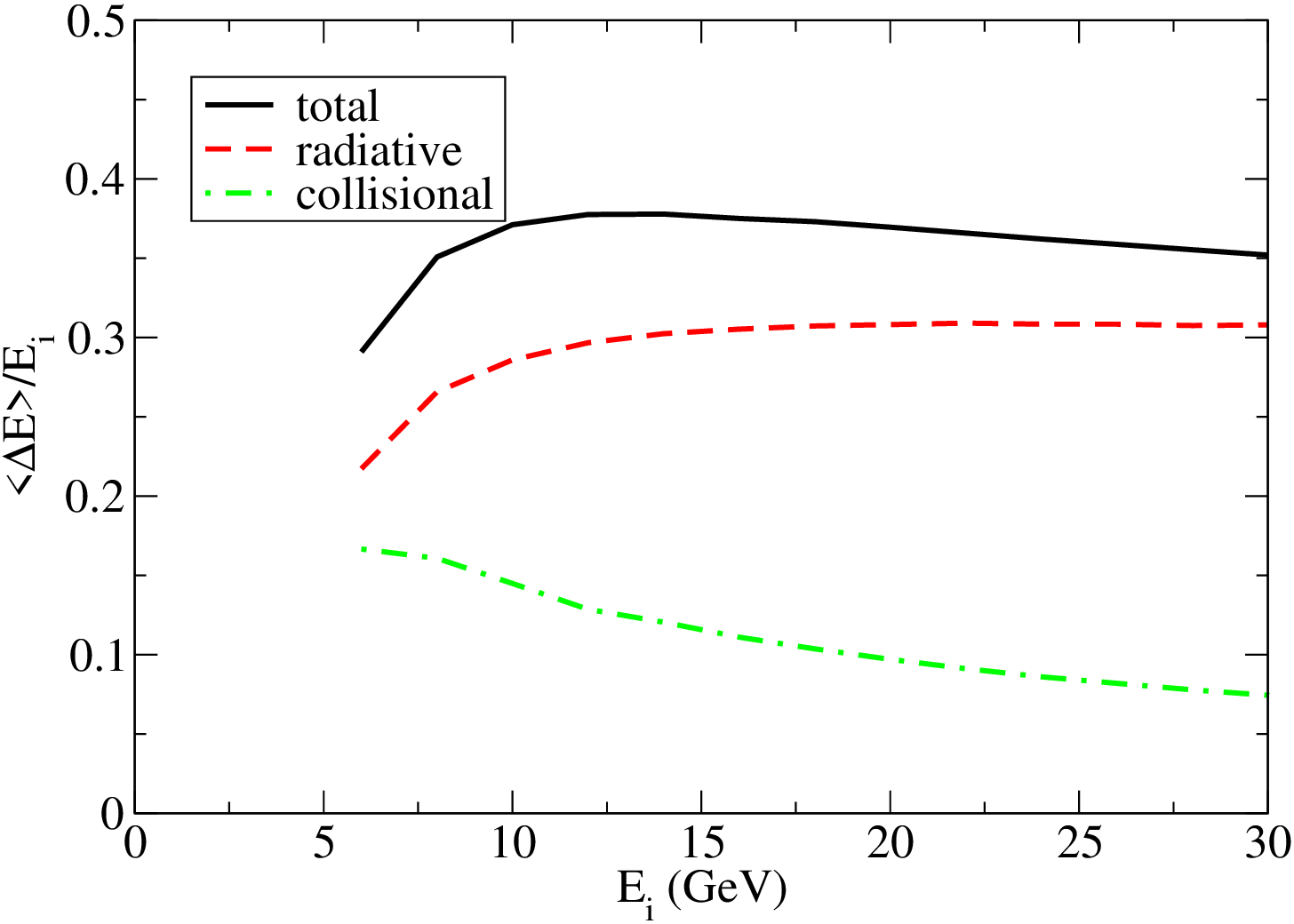}
\end{center}
\caption{(Color online) A single quark jet starts from the center and propagates in plane through the nuclear medium
created in most central Au+Au collisions ($b = 2.4~{\rm fm}$) at RHIC. Left: The final probability distribution $P(E,
t_f)$ of a
single quark with initial energy $E_i = 16~{\rm GeV}$ after passing through the medium. Right (from
\cite{Qin:2007rn}): The fractional averaged energy loss of a quark jet with initial energy $E_i$ after passing through the
medium. } \label{fig12}
\end{figure}

To illustrate how collisional and radiative energy loss influence the time evolution of the leading parton
distributions, we first consider an initial single quark jet starting from center and propagating in plane through the
thermalized medium created in central collisions ($b = 2.4~{\rm fm}$) at RHIC, as shown in Fig. \ref{fig12}. In the
left panel, we compare final parton distribution $P(E,t_f)$ under three different situations: (1) with only
collisional energy loss, (2) with only radiative energy loss, and (3) with both energy loss mechanisms. In the right
panel, the mean energy loss of quark jets after passing through the medium is shown as a function of their initial energy
$E_i$. As expected, induced gluon radiation leads to a much larger mean energy loss than elastic collisions, in
agreement with \cite{Zakharov:2007pj}. While the average energy is not very affected by the addition of collisional
contributions, large differences are observed for the time evolutions of $P(E,t)$ between case (3) and case (2). This
is especially true for energies close to the initial parton energy $E_i$. As the initial parton spectrum in
relativistic nucleus-nucleus collisions is steeply falling, stronger differences in $R_{AA}$ can result.

\begin{figure}[htb]
\begin{center}
\includegraphics[width=7.4cm]{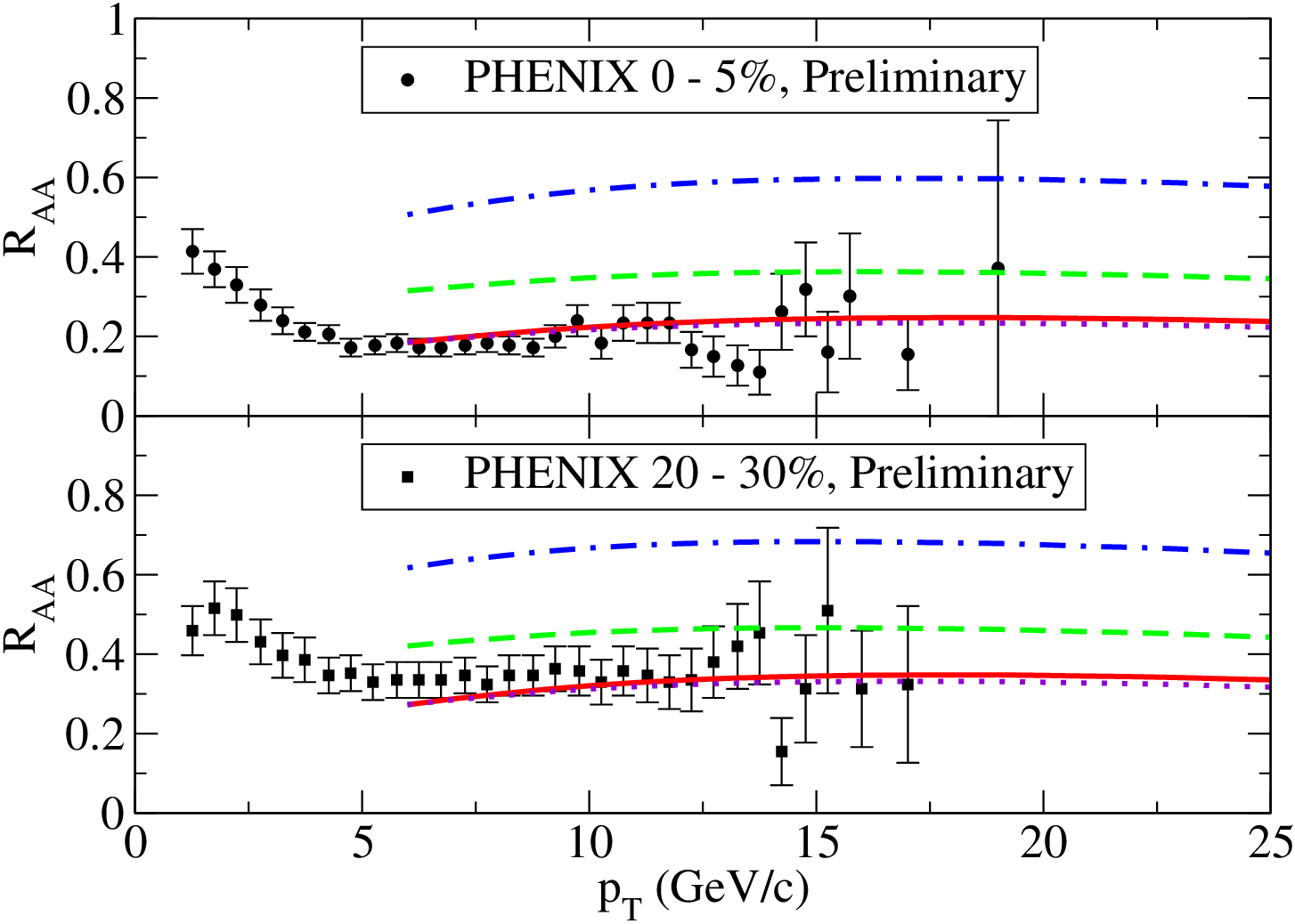}
\includegraphics[width=7.4cm]{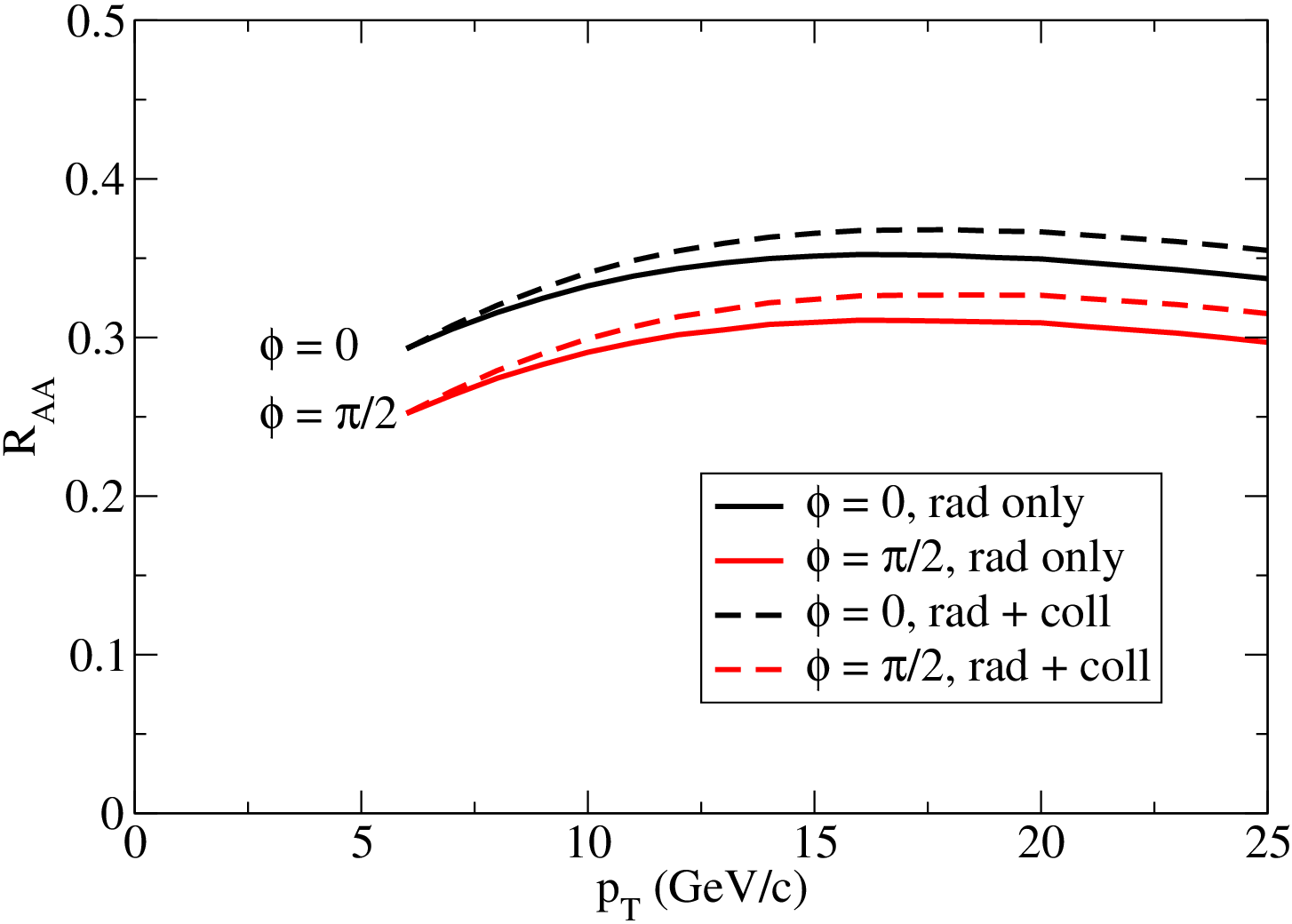}
\end{center}
\caption{(Color online) Left (from \cite{Qin:2007zz, Qin:2007rn}): 
$R_{AA}$ for $\pi^0$ in most central and mid-central
Au+Au collisions at RHIC. The dashed curves includes radiative energy loss only, the dash-dotted curves include
collisional energy loss and the solid curves include both radiative and collisional energy loss. The dotted curves
are the results from \cite{Qin:2007zz} with radiative energy loss only ($\alpha_s = 0.33$). Right: $R_{AA}$ for
$\pi^0$ in plane and out of plane in Au+Au collisions ($b = 7.5~{\rm fm}$) at RHIC before and after the inclusion of
collisional energy loss.} \label{fig34}
\end{figure}

In the left panel of Fig. \ref{fig34}, we present the calculation of $R_{AA}$ for $\pi^0$ measured at mid-rapidity for
two different impact parameters, $2.4$~fm and $7.5$~fm, compared with PHENIX data for the most central (0-5\%) and
mid-central (20-30\%) collisions \cite{Adler:2002xw}. We present $R_{AA}$ for purely collisional (1) and purely
radiative (2) energy loss, as well as the combined case (3). One finds that while the shape of $R_{AA}$ for case (3)
is not strongly different from case (2), the overall magnitude of $R_{AA}$ changes significantly. The magnitude of
$R_{AA}$ is therefore sensitive to the actual form of the parton distribution functions at fragmentation and not only
to the average energy loss of single partons (compare Fig. \ref{fig12}). In \cite{Qin:2007zz}, a systematic study of
the observational implications on $R_{AA}$ measurements due to only radiative energy loss was presented.
Here we recalculate $R_{AA}$ versus reaction plane including elastic collisions, as shown in the right panel of Fig.
\ref{fig34}, and only small differences (after rescaling the coupling strength from $\alpha_s=0.33$ to
$\alpha_s=0.27$) is found in the shape of $R_{AA}$ for the $p_T$ range explored.

\section{Summary}

In conclusion, we have calculated radiative and collisional energy loss of hard partons in the hot
and dense nuclear medium being created at RHIC. The induced gluon bremsstrahlung is treated in the AMY formalism and elastic
collisions are incorporated using a
drag plus diffusion term reproducing average energy loss rate and detailed balance. While the additional
average energy loss
induced by elastic collisions is small in comparison to the radiative one, the time evolutions of the parton distributions
in both cases differ significantly. We find that the inclusion of collisional energy loss significantly
influences the overall magnitude of $R_{AA}$ for $\pi^0$ at RHIC, while the shape of $R_{AA}$
does not show a strong sensitivity. 

\section{Acknowledgments}

We thank C. Nonaka and S. A. Bass for their hydrodynamical calculation \cite{Nonaka:2006yn}. This work was supported
in part by the Natural Sciences and Engineering Research Council of Canada, by the McGill-India Strategic Research
Initiative, and by the Fonds Nature et Technologies of Quebec.

\end{document}